\documentclass[preprint,aps,nofootinbib,showpacs,amsfonts,epsf]{revtex4-2}

\input{epsf.tex}

\usepackage{amsmath}

\newcommand{\E}{{\cal{E}}}
\newcommand{\s}{\sigma}

\newcommand{\bnabla}{\mbox{\boldmath $\nabla$}}

\newcommand{\be}{\begin{equation}}
\newcommand{\ee}{\end{equation}}
\newcommand{\bea}{\begin{eqnarray}}
\newcommand{\eea}{\end{eqnarray}}
\newcommand{\ba}{\begin{array}}
\newcommand{\ea}{\end{array}}
\def\J#1#2#3#4{{#1} {\bf #2}, #3 (#4)}
\def\PRD{Phys. Rev. D}
\def\PR{Phys. Rev.}
\def\PRL{Phys. Rev. Lett.}
\def\PTP{Prog. Theor. Phys.}

\def\APL{Ann. Phys. (Leipzig)}

\def\JMP{J. Math. Phys.}

\def\CQG{Class. Quantum Grav.}

\def\PLB{Phys. Lett. B}

\begin{document}
\draft
\title{Dyonic black holes: \\The theory of two electromagnetic potentials. II}

\author{H. Garc\'ia-Compe\'an,$^\dagger$ V.~S.~Manko,$^\dagger$
and C. J. Ram\'irez-Valdez$^{\dagger\ddagger}$}
\address{$^\dagger$Departamento de F\'\i sica, Centro de Investigaci\'on y
de Estudios Avanzados del IPN, A.P. 14-740, 07000 Ciudad de
M\'exico, Mexico\\$^\ddagger$Theoretical Particle Physics and
Cosmology Group, Department of Physics, King's College London,
University of London, Strand, London, WC2R 2LS, U.K.}

\begin{abstract}
The results obtained in our previous paper are now extended to the
case of stationary axially symmetric dyonic black boles within the
theory of two electromagnetic potentials. We slightly enlarge the
classical Ernst formalism by introducing, with the aid of the $t$-
and $\varphi$-components of the dual potential $B_\mu$, the magnetic
potential $\Phi_m$ which, similar to the known electric potential
$\Phi_e$, also takes constant value on the black hole horizon. We
analyze in detail the case of the dyonic Kerr-Newman black hole and
show how the Komar mass must be evaluated correctly in this
stationary dyonic model. In particular, we rigorously prove the
validity of the standard Tomimatsu mass formula and point out that
attempts to ``improve'' it made in recent years are explained by
misunderstanding of the auxiliary role that singular potentials play
in the description of magnetic charges. Our approach is symmetrical
with respect to electric and magnetic charges and, like in the
static case considered earlier, Dirac strings of all kind are
excluded from the physical picture of the stationary black hole
dyonic spacetimes.
\end{abstract}


\maketitle

\newpage

\section{Introduction}

In the paper \cite{RGM} we have shown that the field of a magnetic
charge is described correctly by the $t$-component of the dual
electromagnetic potential $B_\mu$, so that the semi-infinite
singularities accompanying the $\varphi$-component of the usual
potential $A_\mu$, that must be taken into account during some
mathematical calculations, can not be considered as representing
real physical characteristics of the magnetic charge. In \cite{RGM}
our consideration was restricted to exclusively the static
spherically symmetric dyonic case that ideally suited our objective
of giving simple and clear arguments in favor of our novel approach
to the description of magnetic charges without Dirac strings. In the
present paper we shall expand our analysis to the stationary axially
symmetric dyonic black holes for which the effect of rotation
introduces additional technical difficulties; however, these
difficulties will be circumvented in an elegant way, clearly
confirming the physical conclusions of the previous paper in a more
general situation.

In the next section we shall introduce the nonzero components of the
dual electromagnetic potential $B_\mu$ within the framework of the
well-known Ernst formulation of the stationary axially symmetric
problem \cite{Ern2} and define explicitly the magnetic potential
$\Phi_m$ which, similar to the electric potential $\Phi_e$
introduced long ago by Carter \cite{Car}, takes constant value on
the black-hole horizon. The advantages of the enhanced Ernst
formalism are illustrated here by the example of the dyonic
Kerr-Newman black hole \cite{Car,MGa} for which a complete set of
the corresponding potentials will be constructed. The validity of
the original Tomimatsu mass integral \cite{Tom} will be proven in
Sec.~III with the aid of the symmetrical representation of the
electromagnetic energy-momentum tensor. Discussion of the results
obtained and conclusions can be found in Sec.~IV.

\section{The enhanced Ernst formalism}

In the theory of exact solutions of the Einstein-Maxwell equations,
the Ernst formalism, developed in two papers \cite{Ern1,Ern2} in
1968, occupies an outstanding place as constituting the basis for
various solution generating techniques and different approaches to
the multipole analysis of vacuum and electrovac spacetimes. In
particular, Ernst trivialized the derivation of the Kerr \cite{Ker}
and Kerr-Newman \cite{KNe} black hole solutions that were originally
obtained by means of hardly reproducible procedures.

The main idea of Ernst's formalism is to use the Papapetrou line
element \cite{Pap},
\be d s^2=f^{-1}[e^{2\gamma}(d\rho^2+d z^2)+\rho^2 d\varphi^2]-f(d
t-\omega d\varphi)^2, \label{Papa} \ee
describing a generic stationary axisymmetric electrovac field, with
three arbitrary functions $f$, $\gamma$ and $\omega$ depending on
the cylindrical coordinates ($\rho,z$), for reducing the
corresponding set of the Einstein-Maxwell equations to a fundamental
system of two differential equations for the complex potentials $\E$
and $\Phi$ of the following elegant form:
\bea ({\rm Re}\,{\E}+\Phi\bar\Phi)\Delta{\E}=
(\bnabla{\E}+2\bar\Phi\bnabla\Phi)\bnabla{\E}, \nonumber\\ ({\rm
Re}\,{\E}+\Phi\bar\Phi)\Delta{\Phi}=
(\bnabla{\E}+2\bar\Phi\bnabla\Phi)\bnabla{\Phi}, \label{EE} \eea
where $\Delta$ and $\bnabla$ are the usual three-dimentional
Laplacian and gradient operators, respectively, and a bar over a
symbol means complex conjugation.

The potentials $\E$ and $\Phi$ are related to the metric functions
$f$, $\omega$ and to the $\varphi$ and $t$ components of the
electromagnetic 4-potential $A_\mu=(0,0,A_\varphi,A_t)$ by the
equations
\be \E=f-\Phi\bar\Phi+i\chi, \quad \Phi=-A_t+iA^{'}_\varphi,
\label{EF} \ee
and by the systems of the first-order differential equations
\bea &&\partial_\rho\omega=-\rho f^{-2}[\partial_z\chi+2{\rm Im}
(\bar\Phi\partial_z\Phi)], \nonumber\\
&&\partial_z\omega=\rho f^{-2}[\partial_\rho\chi+2{\rm Im}
(\bar\Phi\partial_\rho\Phi)], \label{wx} \eea
and
\bea &&\partial_\rho A^{'}_\varphi=\rho^{-1}f(\partial_z A_\varphi
+\omega\partial_z A_t), \nonumber\\ &&\partial_z
A^{'}_\varphi=-\rho^{-1}f(\partial_\rho A_\varphi
+\omega\partial_\rho A_t), \label{Afp} \eea
so that the knowledge of $\E$ and $\Phi$ permits one to find the
functions $f$, $\omega$, $A_t$ and $A_\varphi$ from
(\ref{EF})-(\ref{Afp}), while for the determination of the remaining
metric function $\gamma$ one has to solve the system
\bea \partial_{\rho}\gamma&=&\frac{1}{4}\rho
f^{-2}[(\partial_{\rho}{\cal
E}+2\bar\Phi\partial_{\rho}\Phi)(\partial_{\rho}\bar{\cal
E}+2\Phi\partial_{\rho}\bar\Phi)- (\partial_{z}{\cal
E}+2\bar\Phi\partial_{z}\Phi)(\partial_{z}\bar{\cal
E}+2\Phi\partial_{z}\bar\Phi)] \nonumber\\ &&- \rho
f^{-1}(\partial_{\rho}\Phi\partial_{\rho}\bar\Phi
-\partial_{z}\Phi\partial_{z}\bar\Phi), \nonumber\\
\partial_{z}\gamma&=&\frac{1}{2}\rho f^{-2}{\rm Re}[(\partial_{\rho}{\cal
E}+2\bar\Phi\partial_{\rho}\Phi) (\partial_{z}\bar{\cal
E}+2\Phi\partial_{z}\bar\Phi)] - 2\rho f^{-1}{\rm
Re}(\partial_{\rho}\bar\Phi\partial_{z}\Phi), \label{ge} \eea
the integrability condition of which are equations (\ref{EE}).

Note that the potential $A^{'}_\varphi={\rm Im}\,\Phi$ is regarded
in the Ernst formalism as an auxiliary function, the knowledge of
which makes it possible the calculation of the corresponding
magnetic component $A_\varphi$ of the 4-potential $A_\mu$. However,
in our preceding paper \cite{RGM} we have already shown that
$A_\varphi$ does not describe correctly the field of the magnetic
charge, so it seems desirable to supplement the above formalism with
the nonzero components of the dual electromagnetic 4-potential
$B_\mu=(0,0,B_\varphi,B_t)$ that are related to the components $A_t$
and $A_\varphi$ by the first-order differential equations. Indeed,
using the one-form $B=B_t  dt+B_\varphi d\varphi$, we obtain the
desired relations by means of the formula
\be dB=\star F, \label{BF} \ee
where the star denotes the Hodge dual, and $F$ is the usual
electromagnetic 2-form. Taking into account that, on the one hand,
\be dB=d(B_\nu dx^\nu)=\partial_a B_\nu dx^a\wedge dx^\nu,
\label{lBF} \ee
and, on the other hand,
\bea \star F&=&\star d(A_t  dt+A_\varphi d\varphi) \nonumber\\
&=&(g^{t\beta}\partial_a A_t+g^{\varphi\beta}\partial_a A_\varphi)
g^{ab}\sqrt{-g}\,\varepsilon_{b\beta\gamma\delta}dx^\gamma\wedge
dx^\delta, \label{rBF} \eea
($a,b\in\{\rho,z\}$), we get from (\ref{lBF}) and (\ref{rBF}), by
first equating the coefficients at $d\rho\wedge dt$ and $dz\wedge
dt$, the system of differential equations for $B_t$ in terms of
$A_t$ and $A_\varphi$, namely,
\bea &&\partial_\rho B_t=\rho^{-1}f(\partial_z A_\varphi
+\omega\partial_z A_t), \nonumber\\ &&\partial_z
B_t=-\rho^{-1}f(\partial_\rho A_\varphi +\omega\partial_\rho A_t),
\label{BtA} \eea
and then, by equating the coefficients at $d\rho\wedge d\varphi$ and
$dz\wedge d\varphi$, the analogous system for the determination of
$B_\varphi$:
\bea &&\partial_\rho B_\varphi=\rho^{-1}f[(\rho^2
f^{-2}-\omega^2)\partial_z A_t -\omega\partial_z A_\varphi)], \nonumber\\
&&\partial_z B_\varphi=-\rho^{-1}f[(\rho^2
f^{-2}-\omega^2)\partial_\rho A_t -\omega\partial_\rho A_\varphi)].
\label{BfA} \eea

A simple inspection of formulas (\ref{Afp}) and (\ref{BtA}) shows
that the $t$-component of the dual potential $B_\mu$ coincides with
the auxiliary potential $A^{'}_\varphi$ of the Ernst formalism,
i.e.,
\be B_t=A^{'}_\varphi. \label{BAp} \ee
Curiously, it is also possible to identify the component $B_\varphi$
(up to a sign) as the potential $B_2$ introduced in the paper
\cite{Kin} by Kinnersley as part of various matrix potentials of his
solution generating method; in particular, it arises as the
imaginary part of Kinnersley's potential $\Phi_2$.

The knowledge of the full set of the components $A_t$, $A_\varphi$,
$B_t$ and $B_\varphi$ of the 4-potentials $A_\mu$ and $B_\mu$ allows
one to analyze the electric and magnetic fields of the dyonic black
hole solutions in a symmetrical way advocated long ago by Schwinger
\cite{Sch}. For example, as was shown by Carter \cite{Car}, the
electric potential $\Phi_e$ determined as the combination
\be \Phi_e=-A_t-\omega^{-1}A_\varphi, \label{Fe} \ee
assumes constant value on the black-hole horizon. Having introduced
explicitly the dual potential $B_\mu$ into the Ernst formalism, we
can now define ``symmetrically'' the magnetic counterpart of
$\Phi_e$ as
\be \Phi_m=B_t+\omega^{-1}B_\varphi, \label{Fm} \ee
the magnetic potential $\Phi_m$ also taking constant value on the
horizon, which may be considered an important result following from
our approach.

The above said can be well illustrated by the dyonic Kerr-Newman
black hole solution. Following \cite{MGa}, we write its defining
Ernst potentials $\E$ and $\Phi$ in the form
\bea \E&=&\frac{\s x-m-iay}{\s x+m-iay}, \quad \Phi=\frac{q+ip}{\s
x+m-iay}, \nonumber\\ x&=&\frac{1}{2\s}(r_++r_-), \quad
y=\frac{1}{2\s}(r_+-r_-), \quad r_\pm=\sqrt{\rho^2+(z\pm\s)^2},
\nonumber\\ \s&=&\sqrt{m^2-a^2-q^2-p^2}, \label{EFKN} \eea
where the parameters $m$, $a$, $q$ and $p$ stand, respectively, for
the mass, angular momentum per unit mass, electric and magnetic
charges of the black hole (we restrict our consideration to the
real-valued $\s$ only).

The corresponding metric functions $f$, $\gamma$ and $\omega$ have
the form
\bea f&=&\frac{\s^2(x^2-1)-a^2(1-y^2)}{(\s x+m)^2+a^2y^2}, \quad
e^{2\gamma}=\frac{\s^2(x^2-1)-a^2(1-y^2)}{\s^2(x^2-y^2)},
\nonumber\\ \omega&=&-\frac{a(1-y^2)[2m(\s
x+m)-q^2-p^2]}{\s^2(x^2-1)-a^2(1-y^2)}, \label{mfKN} \eea
while for the electric and magnetic components $A_t$ and $A_\varphi$
of the 4-potential $A_\mu$ we have the expressions
\bea A_t&=&-\frac{q(\s x+m)-apy}{(\s x+m)^2+a^2y^2}, \nonumber\\
A_\varphi&=&-py+\frac{a(1-y^2)[q(\s x+m)-apy]}{(\s x+m)^2+a^2y^2},
\label{Atf} \eea
where the integration constant on the right-hand side of $A_\varphi$
has been chosen equal to zero, thus determining the case with two
magnetic ``strings''.

Turning now to the components of the dual 4-potential $B_\mu$, we
see that $B_t$ is obtainable as just the imaginary part of the Ernst
potential $\Phi$, while $B_\varphi$ must be found by solving the
system (\ref{BfA}). The resulting expressions are
\bea B_t&=&\frac{p(\s x+m)+aqy}{(\s x+m)^2+a^2y^2}, \nonumber\\
B_\varphi&=&-qy-\frac{a(1-y^2)[p(\s x+m)+aqy]}{(\s x+m)^2+a^2y^2},
\label{Btf} \eea
where the choice of the integration constant in $B_\varphi$ is the
same as for $A_\varphi$ and defines a pair of electric ``Dirac
strings''.

The only plausible conclusion that can be drawn from the structure
of the components (\ref{Atf}) and (\ref{Btf}) is that the field of
the electric charge $q$ in the dyonic Kerr-Newman solution is
described by $A_t$, and the field of the magnetic charge $p$ is
determined by $B_t$, both components $A_t$ and $B_t$ being well
behaved and asymptotically flat. In turn, the components $A_\varphi$
and $B_\varphi$ possessing the string singularities do not define
the singularity structure of this dyonic black hole solution,
playing exclusively auxiliary mathematical role in some
calculations. For instance, the components $A_\varphi$ and
$B_\varphi$ are needed for the evaluation of the electric  and
magnetic potentials $\Phi_e$ and $\Phi_m$ on the horizon ($\rho=0$,
$-\s<z<\s$, or $x=1$):
\bea \Phi^H_e&=&\left.-A_t-\omega^{-1}A_\varphi\right|_{x=1}=
\frac{q(m+\s)}{(m+\s)^2+a^2}, \nonumber\\
\Phi^H_m&=&\left.B_t+\omega^{-1}B_\varphi\right|_{x=1}=
\frac{p(m+\s)}{(m+\s)^2+a^2}. \label{Femh} \eea

After introducing the angular momentum $J=ma$, and also recalling
that $\omega$ takes constant value on the horizon, so that
\be \omega^{-1}(x=1)\equiv\Omega^H=\frac{a}{(m+\s)^2+a^2},
\label{Om} \ee
one can see that the above formulas verify the Smarr mass relation
\cite{Sma}
\be m=\s+2J\Omega^H+q\Phi^H_e+p\Phi^H_m. \label{Sma} \ee

We now turn to the discussion of the evaluation of the Komar mass
\cite{Kom} in the dyonic Kerr-Newman solution, the issue that also
addresses the question of the distribution of that mass.

\section{Validating Tomimatsu's mass integral formula}

To calculate the Komar \cite{Kom} mass $M$ of a rotating charged
black hole, Tomimatsu \cite{Tom} derived a simple formula
\be M=-\frac{1}{8\pi}\int_H \omega\partial_z\chi\, d\varphi dz,
\label{TMI} \ee
where the integral is taken over the horizon of the black hole.
Formula (\ref{TMI}) was widely used for years in application to
non-isolated black holes in the presence of other black holes or
exterior gravitational fields. In the case of the dyonic Kerr-Newman
black hole, (\ref{TMI}) assumes the form
\be M=-\frac{1}{4}\omega^H[\chi(y=1)-\chi(y=-1)], \label{TKN} \ee
where both $\omega$ and $\chi$ must be taken on the horizon ($x=1$).
It is not difficult to verify that the corresponding $M$ calculated
with the help of (\ref{TKN}) coincides with the mass parameter $m$
in (\ref{EFKN}).

However, the validity of the mass formula (\ref{TMI}) in the
presence of magnetic charge was questioned in the paper \cite{CGa}.
The authors of \cite{CGa} used during their calculations the
conventional representation of the electromagnetic energy-momentum
tensor
\be
T{^\mu}{_\nu}=\frac{1}{4\pi}\left(F^{\mu\alpha}F_{\nu\alpha}-\frac{1}{4}\delta^\mu_\nu
F^{\alpha\beta}F_{\alpha\beta}\right), \label{TE1} \ee
which led them to a specific dyonic configuration with two magnetic
Dirac strings and an additional electromagnetic term in the
integrand of (\ref{TMI}), both strings carrying portions of nonzero
mass, so that the mass parameter $m$ becomes the sum of three
different contributions -- one coming from the surface integral
evaluated on the horizon, and two others arising from the singular
``massive'' Dirac strings. Although the version of the dyonic
Kerr-Newman black hole presented in \cite{CGa} is manifestly
physically inconsistent (see \cite{GMR} for the discussion of
unphysical features of that model), the mathematical computation of
the Komar integral performed in \cite{CGa} looks correct (albeit
some misprints). The explanation of such a seemingly puzzling
situation is quite simple in the framework of the ideas developed in
\cite{RGM} and in the present paper: the pathologies of a specific
representation of the electromagnetic energy-momentum tensor
formally taking part in the calculation of the Komar mass integral
should not be ascribed to the dyonic model itself since the
singularity structure of the magnetic charge is determined by the
well-behaved component $B_t$, and not by the function $A_\varphi$.
In this respect, the desire to automatically associate the
singularities of the auxiliary potentials with the intrinsic
properties of the dyonic black hole would have forced the authors of
\cite{CGa}, after using a different representation of
$T{^\mu}{_\nu}$ involving say the dual electromagnetic tensor
$\tilde F^{\mu\nu}$ only, to draw a new conclusion that it is the
{\it electric} string singularities of the component $B_\varphi$
that contribute to the expression of the Komar mass, with zero
contribution coming from magnetic charge.

As has already been shown in \cite{RGM}, the choice of the
energy-momentum tensor $T{^\mu}{_\nu}$ in the symmetrical
representation
\be T{^\mu}{_\nu}=\frac{1}{8\pi}(F^{\mu\alpha}F_{\nu\alpha}
+\tilde{F}^{\mu\alpha}\tilde{F}_{\nu\alpha}), \label{Tsym} \ee
where
\be F_{\mu\nu}=\partial_\mu A_\nu-\partial_\nu A_\mu, \quad \tilde
F_{\mu\nu}=\partial_\mu B_\nu-\partial_\nu B_\mu,\label{FAB} \ee
permits one to avoid singular sources during the calculation of the
Komar mass integral, reducing the calculational procedure to
exclusively the integrals over the black hole horizon. Although the
paper \cite{RGM} treated the static case, the rotation of the black
hole does not really change the qualitative picture of the
non-rotating model, and below we shall demonstrate that the Komar
mass of the dyonic black hole is obtainable straightforwardly by
means of the original Tomimatsu's mass integral formula (\ref{TMI}),
without the need to consider any singular terms outside the horizon.

In his article \cite{Tom}, Tomimatsu started with the same standard
integral for the calculation of the Komar mass that has been
recently used in the papers \cite{RGM} and \cite{CGa}:
\be M_K=\frac{1}{4\pi}\int_{\infty} D^\nu k^\mu
d\Sigma_{\mu\nu}=\frac{1}{4\pi}\int_{\partial \mathcal{M}} D^\nu
k^\mu d\Sigma_{\mu\nu} +\frac{1}{4\pi}\int_{\mathcal{M}}D_\nu D^\nu
k^\mu dS_\mu, \label{KI} \ee
with the same decomposition into the surface and bulk integrals.

By choosing the horizon of the black hole as $\partial \mathcal{M}$,
Tomimatsu computed the first integral on the right-hand side of
(\ref{KI}) and obtained
\bea \frac{1}{4\pi}\int_{H} D^\nu k^\mu d\Sigma_{\mu\nu}&=&
\frac{1}{8\pi}\int_{H}[-\omega\partial_z\chi+
2\omega{\rm Im}(\Phi\partial_z\bar\Phi)]d\varphi dz \nonumber\\
&=&\frac{1}{8\pi}\int_{H}[-\omega\partial_z\chi+
2\omega(A_t\partial_z B_t-B_t\partial_z A_t)]d\varphi dz,
\label{TI1t} \eea
and he also rewrote the bulk integral on the right-hand side of
(\ref{KI}) in the form
\be \frac{1}{4\pi}\int_\mathcal{M}D_\nu D^\nu k^\mu dS_\mu
=-2\int_\mathcal{M} T{^t}{_t}\sqrt{-g}\,d^3x, \label{TI2t} \ee
and the correctness of formulas (\ref{TI1t}) and (\ref{TI2t}) was
not objected in \cite{CGa}. The authors of \cite{CGa}, however,
questioned Tomimatsu's result of computing the integral on the
right-hand side of (\ref{TI2t}), namely,
\be -2\int_\mathcal{M} T{^t}{_t}\sqrt{-g}\,d^3x=
-\frac{1}{4\pi}\int_{H}\omega{\rm
Im}(\Phi\partial_z\bar\Phi)d\varphi dz, \label{TIr} \ee
which, together with (\ref{TI1t}), gives formula (\ref{TMI}). Though
they rightly pointed out that the representation (\ref{TE1}) of the
energy-momentum tensor used by Tomimatsu requires additionally
taking account of two singular string sources, which modifies the
horizon contribution (\ref{TMI}) of the Komar mass, they still
erroneously ascribed the {\it formal} mass distribution due to
singularities of the auxiliary function to the genuine dyonic
Kerr-Newman space. Actually, we have a strong impression that
Tomimatsu obtained his formula (\ref{TMI}) after deliberately
suppressing the additional electromagnetic term discussed in
\cite{CGa}, with the idea to get a physically consistent expression
for the Komar mass of a black hole. On the other hand, the authors
of \cite{CGa} have restored the additional electromagnetic term in
Tomimatsu's formula (\ref{TMI}) for mathematical consistency, but
this has led them to the physically incorrect result for the mass
distribution in a dyonic black hole.

Remarkably, the validity of Tomimatsu's mass integral (\ref{TMI})
can be readily demonstrated by employing the symmetrical
representation of the electromagnetic energy-momentum tensor
(\ref{Tsym}) for the evaluation of the integral on the right-hand
side of (\ref{TI2t}). Then, following the steps outlined in the
paper \cite{RGM} for that representation, the bulk integral on the
left-hand side of (\ref{TIr}) reduces to the surface integral over
the horizon, yielding
\be -2\int_\mathcal{M} T{^t}{_t}\sqrt{-g}\,d^3x=
\frac{1}{4\pi}\int_{H}(A_t\partial_z B_\varphi-B_t\partial_z
A_\varphi)d\varphi dz, \label{TIc} \ee
where, at the last stage of the computation, we have used the
substitutions
\be \rho^{-1}f[(\rho^2 f^{-2}-\omega^2)\partial_\rho A_t
-\omega\partial_\rho A_\varphi)] =-\partial_z B_\varphi \label{dzB}
\ee
and
\be \rho^{-1}f[(\rho^2 f^{-2}-\omega^2)\partial_\rho B_t
-\omega\partial_\rho B_\varphi)] =\partial_z A_\varphi, \label{dzA}
\ee
the latter relation being the corollary of the first equations of
the systems (\ref{BtA}) and (\ref{BfA}).

Now, combining formulas (\ref{TI1t}) and (\ref{TIc}) in one, and
also taking into account that $\omega$ assumes constant value on the
horizon, we get for the Komar integral (\ref{KI}) the expression
\be M_K= \frac{1}{8\pi}\int_{H}[-\omega\partial_z\chi+ 2\omega
A_t\partial_z(B_t+\omega^{-1}B_\varphi) + 2\omega
B_t\partial_z(-A_t-\omega^{-1}A_\varphi)]d\varphi dz, \label{KMp}
\ee
and lastly, after noting that the second and third terms in the
integrand of (\ref{KMp}) vanish because these contain the
derivatives of the potentials $\Phi_e$ and $\Phi_m$, both potentials
taking constant values on the horizon, we obtain the final
expression for the Komar mass
\be M_K=-\frac{1}{8\pi}\int_H \omega\partial_z\chi\, d\varphi dz,
\label{KMn} \ee
which fully coincides with Tomimatsu's formula (\ref{TMI}).

Therefore, the use of the symmetrical representation (\ref{Tsym}) of
the electromagnetic energy-momentum tensor during the calculation of
the Komar mass integral leads straightforwardly to the original
formula obtained by Tomimatsu in the paper \cite{Tom}. We think this
gives us a nice example of a brilliant physical intuition prevailing
over scholastic mathematical estimates.

\section{Discussion and conclusions}

The derivation of formula (\ref{KMn}) involving exclusively the
integrals over the event horizon unequivocally suggests that the
whole Komar mass evaluated in this way is located inside the horizon
of the black hole. In this respect, it seems remarkable that in the
generic expression (\ref{KI}) for the Komar mass the integration is
set to be performed over a sphere of infinite radius, thus giving an
opportunity to use, if necessary, singular functions during the
computational process. The presence of the electromagnetic field
obviously complicates the evaluation of the Komar mass, both
technically and conceptually, compared to the pure vacuum case
since, as we have seen in our previous paper and in the present one,
the correct choice of the representation of the electromagnetic
energy-momentum tensor is required to avoid the presence of
artificial singularities in the dyonic black holes; consequently, in
the case when an unsymmetrical representation of the energy-momentum
tensor is employed, a very accurate physical interpretation of the
results obtained is needed. Thus, the use of the representation
(\ref{TE1}) in the paper \cite{CGa} urged the authors of that paper
to evaluate the mass integral (\ref{KI}) with the help of the
pathological $\varphi$-component of the potential $A_\mu$. So, it is
not a surprise that they could only arrive, within the framework of
their approach, at the mass distribution spreading along the whole
symmetry axis, and this purely technical result was erroneously
claimed by them an intrinsic property of the dyonic Kerr-Newman
black hole. At the same time, what those authors really did was
simply calculating in a not rational way the same value of the Komar
mass (located entirely inside the black hole horizon) that otherwise
follows directly from Tomimatsu's formula (\ref{TMI}) when the
symmetrical representation of the energy-momentum tensor is used. It
is also clear that since a certain part of the total Komar mass $m$
calculated in the paper \cite{CGa} for the Kerr-Newman dyon comes
from the string singularities, the horizon contribution there
differs from the value obtainable by means of Tomimatsu's formula in
the absence of Dirac strings, which explains the appearance of the
additional electromagnetic term in the mass formula of the paper
\cite{CGa}.

Summarizing the results obtained in our short series of two papers,
it should be first of all pointed out that the knowledge of only the
4-potential $A_\mu$ is generically not sufficient for a correct
description of the electromagnetic field which also requires the
knowledge of the dual 4-potential $B_\mu$. In the case of stationary
axially symmetric fields, these potentials $A_\mu$ and $B_\mu$ have
the nonzero $t$- and $\varphi$-components, namely $A_t$,
$A_\varphi$, $B_t$ and $B_\varphi$, among which it is the
$t$-components $A_t$ and $B_t$ that are the basic key functions
defining the physical properties of the electric and magnetic field,
respectively, in particular their singularity structure, while the
$\varphi$-components $A_\varphi$ and $B_\varphi$ play an auxiliary
role in the description of the electromagnetic field, and the
singularities of the functions $A_\varphi$ and $B_\varphi$ are not
characteristic of the proper electric or magnetic field.

We have shown that the use of a specific representation of the
electromagnetic energy-momentum tensor is able to provoke erroneous
interpretations of the physical properties of dyonic black holes:
thus, the choice of the canonical representation (\ref{TE1}) for
$T{^\mu}{_\nu}$ in the Komar mass integral leads to appearance of
magnetic Dirac strings \cite{Dir} as the sources of mass, while the
representation of $T{^\mu}{_\nu}$ involving only the dual
electromagnetic tensor $\tilde F^{\mu\nu}$ (see formula (8) of
\cite{RGM}) gives rise to massive Dirac strings generated by the
electric charge. This naturally singles out the symmetrical
representation (\ref{Tsym}) of $T{^\mu}{_\nu}$ as the most
appropriate one for the dyonic solutions because no contributions
due to string singularities emerge during the evaluation of the mass
integral with the help of (\ref{Tsym}).

It follows directly from our analysis that Dirac strings (magnetic
and electric ones) must be excluded from the physical picture of
dyonic spacetimes. Nevertheless, the semi-infinite singularities
that are characteristic mathematical attributes of the components
$A_\varphi$ and $B_\varphi$ in the presence of nonzero magnetic and
electric net charges still remain a legitimate part of the general
mathematical toolkit and are expected to be taken into account as
purely mathematical objects in some calculations involving the
functions $A_\varphi$ and $B_\varphi$.

Bearing in mind our basic idea that electromagnetism is necessarily
a theory of two electromagnetic potentials, we have slightly
enlarged the well-known Ernst formalism by explicitly introducing
into it the components $B_t$ and $B_\varphi$ of the dual
electromagnetic potential $B_\mu$. This improves the formalism in
two ways. First, it permits now a unified symmetrical treatment of
the electric and magnetic fields, in particular the introduction for
the first time of the magnetic potential $\Phi_m$ which takes
constant value on the horizon, half a century later than Carter's
electric potential $\Phi_e$ \cite{Car}. Second, after our amendment,
the Ernst formalism looks not only more complete but also
logistically refined: the Ernst auxiliary magnetic function
$A^{'}_\varphi$, which was needed before just for computing the
``genuine'' component $A_\varphi$ of $A_\mu$, and which we
identified as the $t$-component of the dual potential $B_\mu$, now
plays, alongside $A_t$, the leading role in the description of the
electromagnetic field, while $A_\varphi$ plays the role of an
auxiliary function. This, in our opinion, enriches the Ernst
formalism conceptually, as the knowledge of the electromagnetic
Ernst potential $\Phi=-A_t+iB_t$ supplies us directly with the
explicit expressions of the physical components of the
electromagnetic 4-potentials determining the intrinsic properties of
the electromagnetic field, without the need of finding $A_\varphi$.
We notice in this respect that it is the component $B_t$, and not
$A_\varphi$, that takes part for instance in the definition of the
relativistic multipole moments of the electromagnetic field
\cite{Sim,HPe,SAp,FCH,MMeR}, which gives us another good
illustration of a generic secondary role of the component
$A_\varphi$ in the physical analysis.

We hope that our present paper, as well as the paper \cite{RGM},
presenting some new ideas about the description of magnetic charges,
could be also helpful in the search and experimental detection of
the dyonic sources. Of course, a natural expectation would be that
some known elementary particles, in addition to electric charges
they have, might also carry magnetic charges, such particles thus
being the dyonic objects. Taking the dyonic Kerr-Newman solution
considered in Sec.~II as the simplest model for a stationary dyon,
we observe that the corresponding magnetic dipole moment of the
source is $aq$, while the electric dipole moment is equal to $-ap$,
the latter moment arising due to rotation of the magnetic charge.
Therefore, the presence of the electric dipole moment in elementary
particles might be considered in principle as indirect indication
that the particles are endowed with nonzero magnetic charges.

\section*{Acknowledgments}

This work was partially supported by CONACyT of Mexico and by
``Secretar\'ia de Educaci\'on, Ciencia, Tecnolog\'ia e Innovaci\'on
de la Ciudad de M\'exico (SECTEI)'' of the Mexico City.


\begin{references}

\bibitem{RGM} C. J. Ram\'irez-Valdez, H. Garc\'ia-Compe\'an, and
V. S. Manko, Dyonic black holes: The theory of two electromagnetic
potentials, Preceding paper, 2023.

\bibitem{Ern2} F. J.~Ernst, New formulation of the axially
symmetric gravitational field problem. II, \J{\PR}{168}{1415}{1968}.

\bibitem{Car} B. Carter, Black Hole Equilibrium States, in: Black
Holes (eds. C. DeWitt and B. S. DeWitt, Gordon and Breach Science
Publishers, New York, 1973) p. 56.

\bibitem{MGa} V. S. Manko and H. Garc\'ia-Compe\'an, Smarr formula for
black holes endowed with both electric and magnetic charges,
\J{\CQG}{35}{064001}{2018}; arXiv:1506.03870 [gr-qc].

\bibitem{Tom} A. Tomimatsu, Equilibrium of two rotating charged
black holes and the Dirac string, \J{\PTP}{72}{73}{1984}.

\bibitem{Ern1} F. J.~Ernst, New formulation of the axially
symmetric gravitational field problem, \J{\PR}{167}{1175}{1968}.

\bibitem{Ker} R. P. Kerr, Gravitational Field of a Spinning
Mass as an Example of Algebraically Special Metrics,
\J{\PRL}{11}{237}{1963}.

\bibitem{KNe} E. Newman, E. Couch, K. Chinnapared, A. Exton,
A. Prakash, and R. Torrence, Metric of a rotating charged mass,
\J{\JMP}{6}{918}{1965}.

\bibitem{Pap} A. Papapetrou, Eine rotationssymetrische L\"osung in
der allgemeinen Relativit\"atstheorie, \J{\APL}{12}{309}{1953}.

\bibitem{Kin} W. Kinnersley, Symmetries of the stationary
Einstein-Maxwell field equations. I, \J{\JMP}{18}{1529}{1977}.

\bibitem{Sch} J. Schwinger, Magnetic charge and the charge quantization condition,
\J{\PRD}{12}{3105}{1975}.

\bibitem{Sma} L.~Smarr, Mass formula for Kerr black holes,
\J{\PRL}{30}{71}{1973}.

\bibitem{Kom} A.~Komar, Covariant conservation laws in general
relativity, \J{\PR}{113}{934}{1959}.

\bibitem{CGa} G. Cl\'ement and D. Gal'tsov, On the Smarr formula
for rotating dyonic black holes, \J{\PLB}{773}{290}{2017}.

\bibitem{GMR} H. Garc\'ia-Compe\'an, V. S. Manko, and E. Ruiz,
Comments on two papers of Cl\'ement and Gal'tsov, arXiv:2006.00793
[gr-qc].

\bibitem{Dir} P. A. M. Dirac, The theory of magnetic poles,
\J{\PR}{74}{217}{1948}.

\bibitem{Sim} W. Simon, The multipole expansion of stationary
Einstein-Maxwell fields, \J{\JMP}{25}{1035}{1984}.

\bibitem{HPe} C. Hoenselaers and Z. Perj\'es, Multipole moments
of axisymmetric electrovacuum spacetimes, \J{\CQG}{7}{1819}{1990}.

\bibitem{SAp} T. P. Sotiriou and T. A. Apostolatos, Corrections
and comments on the multipole moments of axisymmetric electrovacuum
spacetimes, \J{\CQG}{21}{5727}{2004}.

\bibitem{FCH} G. Fodor, E. S. Costa Filho, B. Hartmann, Calculation of multipole
moments of axistationary electrovacuum spacetimes,
\J{\PRD}{104}{064012}{2021}.

\bibitem{MMeR} V.~S.~Manko, I. M. Mej\'ia, E. Ruiz, Metric of
a rotating charged magnetized sphere, \J{\PLB}{803}{135286}{2020}.


\end{references}
\end{document}